# Effects of substituting calcium for yttrium on the superconducting properties of YBa$_2$Cu$_3$O$_z$ bulk samples

E. K. Nazarova[1, 2], A. J. Zaleski[3], A. L. Zahariev[1], A. K. Stoyanova-Ivanova[1], K. N. Zalamova[1]

[1]*Institute of Solid State Physics, BAS, Sofia, Bulgaria*
[2]*International Laboratory of High Magnetic Fields and Low Temperature, 53-421, Wroclaw, Poland*
[3]*Institute of Low Temperature and Structure Research, PAS, 50-950 Wroclaw, Poland*

**Abstract**

We report systematic studies of AC magnetic susceptibility and transport properties of Y$_{1-x}$Ca$_x$Ba$_2$Cu$_3$O$_z$ bulk samples with 0≤$x$≤0.4. Single phase materials, reduction of carrier concentration and decrease of $T_c$ to 83K were obtained at doping levels up to 20%. For Y$_{0.7}$Ca$_{0.3}$Ba$_2$Cu$_3$O$_z$ sample the improvement of grains boundary transport and screening capabilities has been observed as a result of the optimal ratio between carrier concentration and impurity phase BaCuO$_2$ presence. The appearance of bulk pinning and nonlinear effects starting at the highest temperature were detected also.

**Introduction**

The solution of the problem of critical current limitation in HTSC by weak links is crucial for the large-scale application of these materials. To overcome this restriction in YBCO system different methods have been proposed. They are connected with increasing the number of current carriers by hole doping [1-4], improving grain alignment [5, 6] and maximization of the effective grain-boundary cross-section [7]. G. Hammerl et al. offered to use all these three methods together when a coated YBCO tape was fabricated [8].

Substitution of $Y^{3+}$ ion by $Ca^{2+}$ ion with similar ionic radius but lower valence is one of the methods for increasing the hole concentration in YBCO system. By using Ca- substitution in YBCO thin films A. Schmehl et.al. [1] obtained the critical current density exceeding the highest value (by more than a factor of seven) ever reported for the no substituted systems. It was observed for the grain boundary misorientation angle of 24 degrees. The grain boundary behavior, which prevents high-field applications, is the essence of HTS thin-film electronics technologies. For successful fabrication of Josephson junction with $Y_{1-x}Ca_xBa_2Cu_3O_z$ interlayer a good understanding of this hole doped material is needed also [9].

The effect of Ca hole doping is experimentally observed as an insulator-to-metal transition in the system $Y_{1-x}Ca_xBa_2Cu_3O_{6.1}$ [2, 10]. The hole doping increases the total density of states, as it has been shown theoretically, moves the Fermi energy, $E_F$, inside the valence band and raises significantly $T_c$ [11, 12].

From neutron diffraction refinement [13], the occupancy of Ca on the Y site was found to be close to 100% up to $x$=0.15-0.18. For higher values of $x$, Ca distributes over Y- and Ba- sites in a ratio of 3:1 [12] but as $x$ tends towards 0.5 the distribution moves towards 1:1, prerequisite impurity phase formation [14, 15].

The previous studies imply that the reduction of charge at the Y site due to Ca substitution is balanced by oxygen loss [15-19] as a result hole concentration little exceeds that of non substituted compound [3]. The corresponding two plateaus in critical temperature vs. oxygen content dependence still occur and are shifted toward lower oxygen concentration than that in pure YBCO system [19]. Finally, when the oxygen content decreases the value of lattice constant $c$ increases and the distance between $CuO_2$ planes decreases. This enhances the antiferromagnetic correlation between Cu-spins in the planes along the $c$-axis and destroys the superconductivity. The shorter distance between $CuO_2$ planes - the lower $T_c$ is obtained. This tendency is kept also in the case of Ca substituted YBCO compounds [19].

The influence of Ca doping on the transport properties of bulk YBCO samples is also important and was discussed earlier [3, 4, 18]. In this work we provide a systematic study of magnetic and transport properties of bulk $Y_{1-x}Ca_xBa_2Cu_3O_z$ ($x = 0 - 0.4$) samples. Although these specimens consist of network of randomly oriented grains, it is important to emphasize that Ca plays much more interesting role than just a carrier source.

**Experimental**

*Sample preparation*: $Y_{1-x}Ca_xBa_2Cu_3O_z$ samples were prepared from the high purity $BaCO_3$, $Y_2O_3$, $CuO$ and $CaCO_3$ powders. The obtained mixture was ground and three times sintered by the standard solid state reaction. The first sintering was at 900°C in the flowing oxygen for 21 hours. After grinding the powder was sintered at 930°C for the second time at the same atmosphere followed by slow cooling and additional annealing at 450°C for 2 hours. Tablets were pressed, sintered for the third time at 950°C for 23 hours and subsequently annealed at 450°C for 23 hours. The density of monophase Ca substituted samples is approximately 82% of the theoretical value and for non-substituted sample is a little lower, about 80%.

*Sample investigations:* Standard X-ray powder diffraction analysis with Co(K$\alpha$) radiation was used for examination of material structure. Magnetic AC susceptibility was measured with commercial MagLab - Oxford 7000 susceptometer in the AC-fields of 0.1, 1 and 10 Oe at 1000 Hz. For these measurements the samples were cut into bars of approximately 2.5x2.5x15 mm$^3$. The demagnetization factor $N$ for the all samples was estimated roughly to be about 0.155 (SI units).

Temperature dependence of critical current density at zero magnetic field was measured in specially designed device from 77K up to $T_c$ using four probe contact method. The necessary temperature was stabilized by scheme including PID controller with 10mK accuracy [20]. The current in the circuit was determined by measuring voltage on 10Ω standard resistor. The 10 μV/cm voltage criterion was used for critical current definition. Voltages have been measured by digital voltmeters with sensitivity 100nV and accuracy 0.5μV. The investigated samples have been prepared with the minimal cross-section 0.8 x 2 mm$^2$, in order to avoid heating of the samples, and 15 mm length.

**Results and Discussion**

It was established from the X-ray powder diffraction measurements that all the prepared samples were orthorhombic. The characteristic orthorhombic splitting due to the reflection from (006), (020), and (200) planes is presented in Fig.1 (2θ = 54 ÷58 degrees). By detailed analysis in the range of 2θ = 30÷40 degrees, where the main peaks of $BaCuO_2$ phase occur, it was shown that detectable amount of this impurity phase appeared only in samples with $x > 0.2$. For clear presentation of these small peaks in Fig.1, x-axis is interrupted and only narrow region of 2θ = 32 ÷36 degrees is shown.

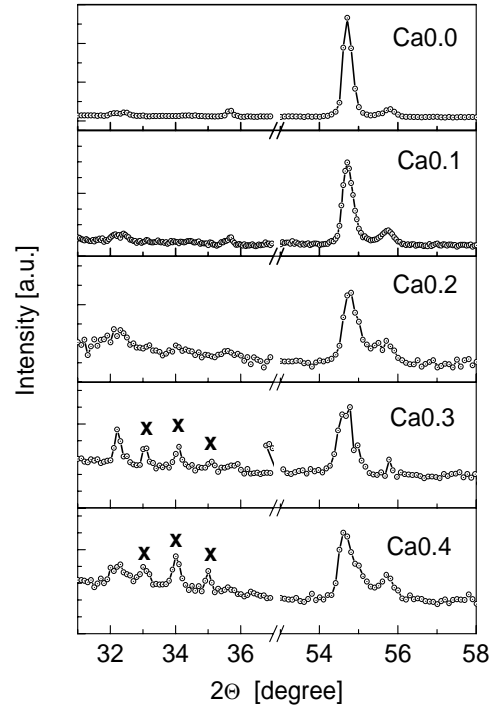

Fig.1. XRD patterns between 2θ=32-36 and 54-58 degrees using Coκα radiation for investigated samples. The symbol (X) indicates main peaks of $BaCuO_2$ phase.

This XRD result was supported by scanning electron microscopy study, which did not reveal $BaCuO_2$ phase for the sample with $x=0.2$, but $BaCuO_2$ grains with dimensions comparable to the grains of basic YBCO phase (10μm) has been found in the sample with $x=0.3$.

This is in agreement with the observations of other authors [13]. The appearance of $BaCuO_2$ phase indicates that Ca starts to substitute not only for Y but for Ba also. These substitutions (Ca for Y and Ca for Ba) are possible in different extent in our samples. In the specimens with $x \leq 0.2$ only Ca for Y substitution takes place since $BaCuO_2$ is not found. For $x > 0.2$, due to the substitution of Ca for Ba and $BaCuO_2$ formation, Cu vacancies and vacancies on Y position appeared. According to P. Starowicz et al. [10] about 8.9% Cu atoms are incorporated in $BaCuO_2$ when $Y_{0.7}Ca_{0.3}Ba_2Cu_3O_{6.12}$ compound is synthesed. We suppose that Ca substitution for Ba creates defects in the unit cell but do not destroy it totally. If it is not the case Y and Ca have to be indicated in the XRD (especially in the sample with $x=0.4$) but we did not see them. With increasing Ca content (for $x=0.4$) the number of Cu vacancies become larger than these in the sample with $x=0.25$ as the Ca substitution on Y and Ba position is moved from 3:1 towards 1:1 [14].

The oxygen concentration was experimentally determined by spectrophotometric method [21]. The reproducibility of oxygen measurements was checked for two samples with $x=0.1$. The obtained results for oxygen content, $z$ agree within $\Delta z=0.05$. By increasing the statistics $\Delta z$ can be lowered down to 0.01. A comparison has been made between our data and the results of the other authors using similar sintering procedure (Fig.2). The largest deviation of the data is observed for sample with $x=0.3$ due to the second phase presence. For the other $x$ concentration the agreement is satisfactory. The equilibrium oxygen content $z=7-x/2$ for different Ca concentrations [3] is also presented.

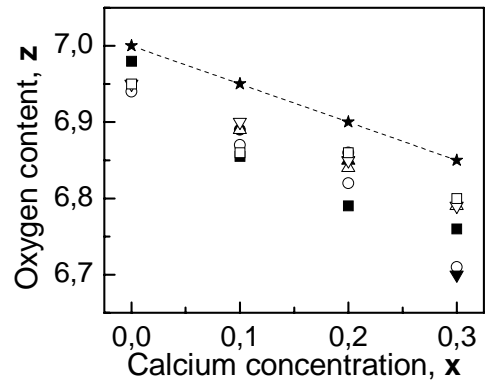

Fig.2. Oxygen content, $z$ vs. calcium concentration, $x$ : ■ -our data, ○ - A. Manthiram et al. (Ref. 15), ● - C. Greaves et al. (Ref. 16), △- A. Manthiram et al. (Ref.17), ▲- G. Xiao et al. (Ref. 18), ▽- B. Fisher et al. (Ref.23), ▼- C. Legros-Gledel et al. (Ref.2), □- Y. Tokura et al. (Ref.22), ★- $z=7-x/2$.

On Fig.3 real parts of the AC magnetic susceptibility as a function of temperature are presented for all investigated samples, for $H_a$= 0.1 Oe and $f$ = 1000 Hz. From this figure the $T_{c,\,onset}$ for intra- ($T_c^{inta}$) and intergranualar ($T_c^{inter}$) screening can be evaluated for all specimens. The Ca free sample Ca0.0 has the highest oxygen concentration 6.98 and the narrowest temperature transition region for screening (90-91K).

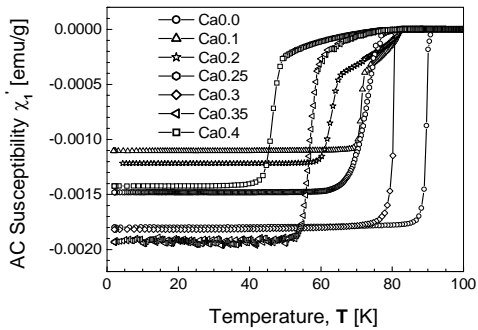

Fig.3. The temperature dependences of a real part fundamental AC magnetic susceptibility for all investigated samples at $H_{AC}$=0.1 Oe and $f$=1000 Hz.

These parameters are characteristic for the material within first plateu in the well known dependence of $T_c$ vs. oxygen content reported by Cava et.al. [24]. With Ca substitution for Y the critical temperature of the samples decreases. Some authors claim that the depression of $T_c$ with $x$ is caused by the decrease of oxygen content in the chains [19], other regard that trapping of mobile holes or some mechanisms connected with oxygen vacancies disorder are responsible for such behavior [2, 17, 18, 23].

From our investigation it turned out that oxygen concentration in samples Ca0.1 and Ca0.2 decreases more distinctly than resulting from $z=7-x/2$. Every lost oxygen atom lead to reduction of hole content with one hole per $Cu(2)O_2$ plane, while every $Y^{3+}$ ion substituted by $Ca^{2+}$ increases carrier concentration with ½ hole per plane. In fact, additional carriers introduced by Ca were compensated by oxygen reduction [3, 15, 25]. $T_c^{inta}$ and $T_c^{inter}$ decreased with growth of $x$.

In all samples with $x>0.2$, $BaCuO_2$ phase exists due to the Ca substitution on Ba place. This results in formation of vacancies in Y position and Cu vacancies as well. One vacancy on Y position decreases the positive charge in the unit cell and from the considerations of charge balance enlarges the hole concentration by 1.5 per plane. The effect of Cu vacancies is identical. This increase carrier concentration although the oxygen decrease and as it is seen from Fig.3 samples with $x>0.2$ posses better screening capabilities than those with $x\leq0.2$. Also sample Ca0.3 screened the applied field ($H_{ac}$=0.1 Oe) by only intergranular currents.

Further increasing of Ca concentration in sample Ca0.4 caused depression of intergranular current due to the increase of insulating $BaCuO_2$ phase distributed between grains. The other nondesirable effect is growing the number of Cu vacancies, which destroy seriously grain structure and together with oxygen deficit caused the largest intragrain screening down to 50K.

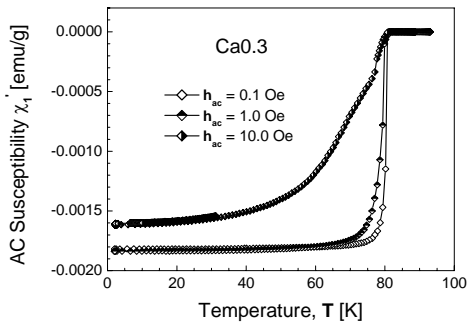

Fig.4. The temperature dependences of a real part fundamental AC magnetic susceptibility for Ca0.3 sample at different $H_{AC}$ amplitudes and $f$=1000 Hz.

On Fig.4 real parts of fundamental AC magnetic susceptibility as a function of temperature are presented for Ca0.3 sample at different AC magnetic field amplitudes. For $H_{AC}$=0.1Oe and 1Oe sample is screened by intergranular current only. By increasing the magnetic field amplitude up to 10Oe two temperature ranges with intra- and intergranular screening appear. In all other substituted samples intra- and intergranular screening is observed even at $H_{AC}$=0.1Oe. In specimen Ca0.3 the higher fields has been screened by the higher currents supported by the large carrier concentration in this sample.

Fundamental susceptibility carries information about shielding and energy dissipation, while higher harmonics have information about the pinning mechanism of vortices. The third harmonic let one not only to discriminate between the linear and nonlinear AC response, but also yields a very different polar plot depending on the type of screening and proposed critical state model [26, 27], which is experimentaly observed also [28, 29].

On Fig.5 (a, b, c) $\chi_3$ vs. $T$ dependencies for Ca0.2, Ca0.3 and Ca0.35 samples are presented. In Ca0.2 sample $\chi'_3$ and $\chi''_3$ signals are essentially positive and suggest the absence of bulk pinning. Similar dependencies have been found for Ca0.1 sample. Absolutely different dependences were obtained for Ca0.3 and Ca0.35 samples, where both signals $\chi'_3$ and $\chi''_3$ are negative, pointing out the existence of good bulk pinning.

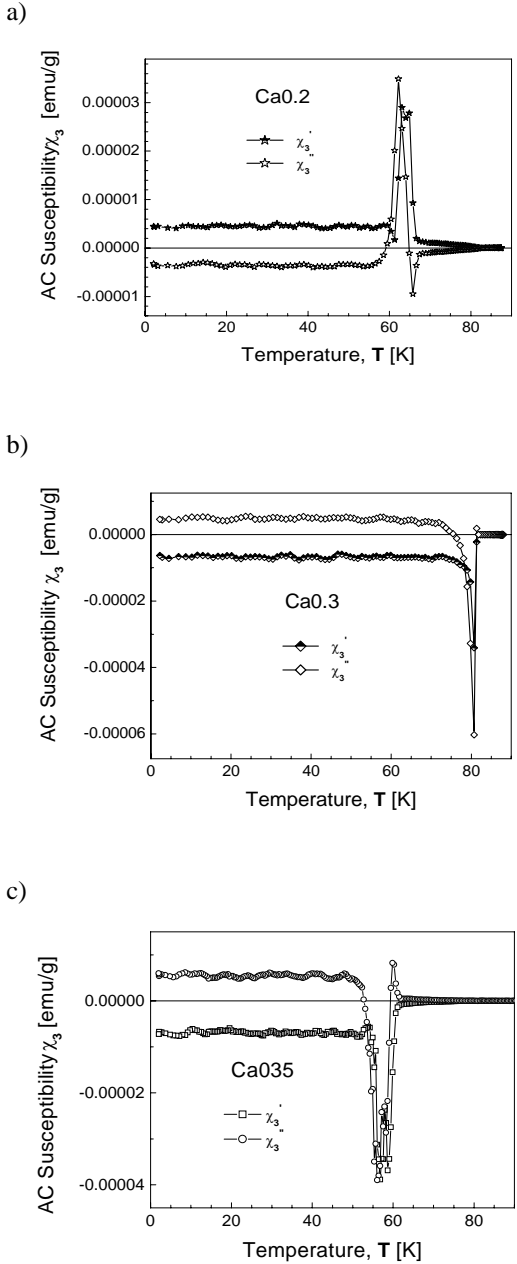

Fig.5. The temperature dependences of $\chi'_3$ and $\chi''_3$ at $H_{AC}$=0.1 Oe and $f$=1000 Hz for a) Ca0.2 b) Ca0.3 and c) Ca0.35 samples.

Vacancies on Y position, which are point defects generated by chemical substitution, can be possible pinning centres. The effect of point defects as a pinning centers was reported by Harada *et. al.* [30, 31]. Determination of the exact nature of pinning centers in these samples needs further work. It seems that in Ca0.3 sample the best correlation is obtained between the carrier concentration, pinning centers and impurity phase. Further increasing of Ca content increases the amount of impurity phase and the number of Cu vacancies as well, which diminishes the carrier concentration and deteriorate superconducting current.

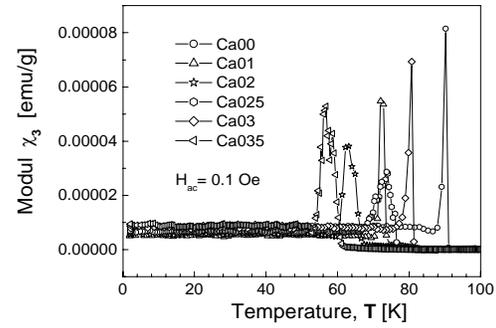

Fig.6. The temperature dependences of $|\chi_3|$ for all indicated samples at $H_{AC}$=0.1Oe and $f$=1000 Hz.

The bulk pinning hysteresis losses are proportional to the area of hysteresis loop, which in turn is proportional to the imaginary part of complex permeability (or susceptibility) [32]. Thus $\chi''_1$ can be used for determination of AC losses per cycle, per unit volume, while the modulus of the third harmonic, $|\chi_3|$ detected precisely the onset of nonlinear effects [33, 34]. Higher harmonic response is useful also for separating and evaluating the hysteresis loss due to the flux pinning in a superconductor sheated with metals [35, 36]. On Fig.6 $|\chi_3|$ vs. temperature are presented for all samples. In consistency with previous results the nonlinearity started at the highest temperature in Ca0.3 sample.

Fig.7 shows the typical field dependence of the real part of fundamental AC susceptibility at various temperatures for the Ca substituted samples. The temperatures have been choosen to cover the region of intergranular transition in all samples.

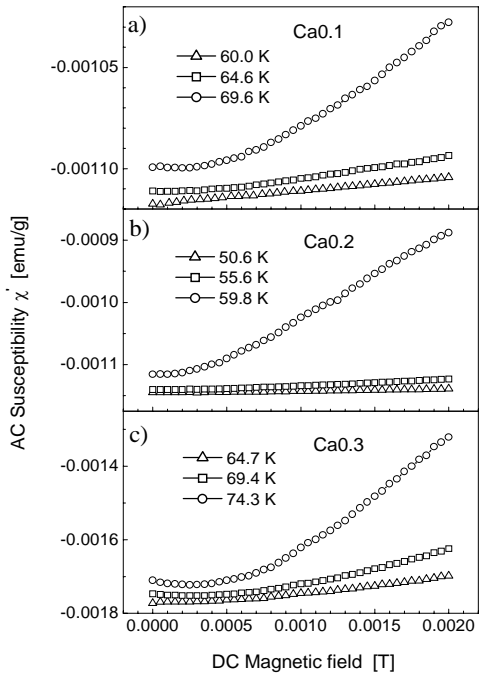

Fig.7. Magnetic field dependences of intergranular $\chi'_1$ at different temperatures for samples a) Ca0.1, b) Ca0.2 and c) Ca0.3 at $H_{AC}$=0.1 Oe; $f$=1000 Hz.

It is seen from the Fig.6 that the intergrain magnetic field penetration starts at the highest temperature and magnetic field value in Ca0.3 specimen. Decreasing of screening effect in this sample at $T$=74.3K may result from two different mechanisms. The first one is a current supression due to the increasing the magnetic field and the second one is connected with the thermaly activated flux flow (TAFF). At temperatures below $T_c$, as the DC field increases, the TAFF induced resistance will increase and thus the shielding effect of superconductor decreases. This mechanism seems to be realistic since the non zero $|\chi_3|$ signal was found at $T \approx 78$ K (Fig.6) when only AC field was applied. Independently, which of both mechanisms is active, the observed decrease in $\chi_1'$ ($H$) at the highest $T$ and DC magnetic field values for the Ca0.3 specimen is in agreement with the fact that this sample possess the optimal carrier concentration and pinning. Transport measurements also support this finding. On Fig.8 critical current density, $J_c$, as a function of ($T_c$-$T$) has been presented for the two samples with the highest $J_c$ and $T_c$ values.

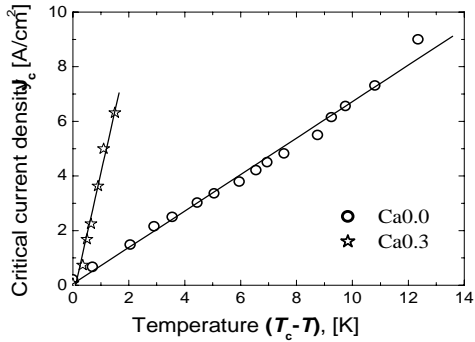

Fig.8. $J_c$ vs. ($T_c$-$T$) for the Ca free sample and Ca0.3 speciment. The lines present the best linear fit to data.

Other samples have $T_c^{inter} < 77K$ and we are not able to measure them. The lines on Fig.8 present the best linear fit to the data. It is known that high temperature superconducting bulk samples may be treated as a system of superconductor- insulator- superconductor (S-I-S) Josephson junctions for which $J_c$ is linear function of temperature for $T<T_c$ [37, 38]. In agreement with previously presented magnetic measurements Ca0.3 sample shows higher slop of the plot of $J_c$ vs. ($T_c$-$T$) than Ca free sample indicating higher critical current density at temperatures below $T_c$.

## Conclusions

In conclusion we have demonstrated that in $YBa_2Cu_3O_z$ samples intergranular critical current may be increased without grain alignment or maximization of grain boundary area by solely substitution of yttrium with calcium. The best result has been found for the 30% Ca substituted sample, where Ca acts not only as a carrier provider but generates vacancies on Y and Cu positions. This decreases the positive charge in the unit cell and in fact raises the carrier concentration.

The second effect of Ca substitution, which has been detected by higher harmonics of AC magnetic susceptibility measurments, is the appearance of bulk pinning in samples with $x>0.2$. It can be related to the point defects generated by the substitution. We establish that in our investigated sample series the optimal ratio between carrier concentration, number or strength of pinning centers and an impurity phase is obtained in $Y_{0.7}Ca_{0.3}Ba_2Cu_3O_z$.

## Àcknowledgements

Authors are grateful to K. Nierzewski for SEM investigations, D. Kovacheva, K. Petrov, T. Nedelcheva